\documentclass[12pt]{article}
\begin{document}

\title{The Wave Functions for the Free-Fermion Part of the Spectrum of 
the $SU_q(N)$  Quantum Spin Models}
\author{F.~C.~Alcaraz$^{\rm a}$, Yu.~G.~Stroganov$^{\rm a,b}$\\
\small \it $^{\rm a}$Universidade de S\~ao Paulo, Instituto de F\'{\i}sica de S\~ao
Carlos,\\[-.5em]
\small \it C.P. 369, 13560-590, S\~ao Carlos, SP Brasil\\
\small \it $^{\rm b}$Institute for High Energy Physics\\[-.5em]
\small \it 142284 Protvino, Moscow region, Russia}
\date{}

\maketitle

\begin{abstract}
We conjecture that the free-fermion part of the eigenspectrum observed recently for the $SU_q(N)$
Perk-Schultz spin chain Hamiltonian in a finite lattice with $q=\exp (i\pi (N-1)/N)$ is a
consequence of the  existence of a special simple eigenvalue  
 for the transfer matrix of the auxiliary inhomogeneous $SU_q(N-1)$ vertex
 model which appears in the nested Bethe ansatz approach. We prove that this
 conjecture
 is valid for the case of the $SU(3)$ spin chain with  periodic boundary
 condition. In this case  we
 obtain a formula for the  components of the eigenvector of the auxiliary
 inhomogeneous 6-vertex model ($q=\exp (2 i \pi/3)$), 
which permit us to find one by one all components of this 
eigenvector and consequently to find the eigenvectors of the 
free-fermion part of the eigenspectrum of the $SU(3)$ spin chain.
Similarly as in the  known case of the $SU_q(2)$ case at $q=\exp(i2\pi/3)$ our numerical 
and analytical studies induce some conjectures for special rates of 
correlation functions. 
\end{abstract}

\vskip 1em

\begin{center}{\bf1. Introduction}
\end{center}

It was found very recently that  part of the eigenspectrum of some quantum spin models is given by
a  sum of free-fermion quasienergies \cite{AS1,AS2}. In particular, the ground state energy of the
$SU_q(N)$
invariant Perk-Schultz Hamiltonian
 \begin{eqnarray}
\label{H}
&&H_{q} =\sum_{j=1}^{L-1} H_{j,j+1} 
\end{eqnarray}
where   
\begin{eqnarray}
&&H_{i,j} =-\sum_{a=0}^{N-1} \sum_{b=a+1}^{N-1} 
( E_i^{ab} E_{j}^{ba} + E_i^{ba} E_{j}^{ab}- q E_i^{aa} E_{j}^{bb} 
- 1/q  E_i^{bb} E_j^{aa} ) \nonumber
\end{eqnarray}
and $E^{ab}$ are $N\times N$ matrices with elements 
$(E^{ab})_{cd} = \delta_c^a \delta_d^b$,
is given by 
\begin{equation}
\label{Eminfree}
E_0 = 1 + 2 (1-L+n) \cos(\frac{\pi}{N}) -\frac{\sin \pi (2n+1)/2L}{\sin
\pi/2L}
\end{equation}
for the special value of the deformation parameter
\begin{equation}
\label{q}
q=\exp \biggl (\frac{i \pi (N-1)}{N} \biggr).
\end{equation}
The lattice size defining the quantum chain is $L$ and the 
parameter $n$ in (\ref{Eminfree})
is given by  the integer part of 
$L/N$.

Moreover  an amazingly simple formula was found (conjectured) for the ground state energy of
$SU_q(N-1)$ Hamiltonian for the same value of deformation parameter

 \begin{equation}
\label{Eminpred}
\tilde {E}_0 = - 2 (L-1) \cos(\frac{\pi}{N}).
\end{equation}

While  for the $SU_q(N)$ model  many eigenenergies  can be described as the sum of free-fermion
quasienergies, in the case of 
the $SU_q(N-1)$ there exists a single energy level in this class, 
namely the ground
state energy given in (\ref{Eminpred}). Actually this state can be 
 considered as a special state of the $SU_q(N)$ model 
in the sector where only ($N-1$) classes of particles are present
(see conjecture 3 of \cite{AS2}).
We intend to show  in this paper that the corresponding wave function 
possess nice combinatorial properties and its components  play a very 
important role in the nested Bethe ansatz approach being used for the 
 construction of all other free-fermion eigenstates. This comes from 
the fact that  using the nested bethe ansatz (NBA) method for 
the 
$SU_q(N)$ invariant Hamiltonian (\ref{H}) we obtain (see, for example, \cite{freeBAE})
an auxiliary transfer matrix of the inhomogeneous $SU_q(N-1)$ invariant vertex model. We
conjecture that this matrix has an unique factorizable eigenvalue
that  reduces some of the NBA equations into a very simple form 
leading to the free-fermion-like structure. 
As a result we obtain the free-fermion part of the eigenspectrum of the 
Hamiltonian (\ref{H}). All related eigenvectors can be found 
from the knowledge of this  unique eigenvector 
of the transfer matrix of the inhomogeneous 
$SU_q(N-1)$ vertex model.
In the  homogeneous case the transfer matrix  commutes with the 
$SU_q(N-1)$ invariant Hamiltonian and this special eigenvector  reduces 
to the ground sate eigenvector with energy (\ref{Eminpred}).

As far as a periodic boundary case is concerned, we have found the free-fermion spectrum only for
the $SU(3)$ Perk-Schultz model with 
$q=\exp (2 i\pi/3)$ (we do not consider here the free-fermion point of the $SU(2)$ model, or
equivalently the standard  $XY$ spin chain).
We show below that the free-fermion part of the eigenspectrum of the $SU(3)$ 
model with
periodic boundary condition can be explained by the existence of an unique factorizable eigenvalue
for the transfer matrix of the inhomogeneous 6-vertex model with
$q=\exp (2 i \pi/3)$. We prove the existence of this eigenvalue using different methods described
in papers \cite{AS2},\cite{Snew}.
Comparison of these approaches  sheads a new light on recently discovered combinatorial properties
of the ground state wave function for the odd length 
XXZ spin model with ($\Delta=-1/2$) 
\cite{alc1},\cite{YGS},\cite{ALL} (see also \cite{XYZ} 
for similar results for the XYZ spin chain).

\vskip 1em

\begin{center}{\bf2. The $SU_q(N)$ invariant model}
\end{center}

The Hamiltonian (\ref{H}) describes the dynamics of a system containing 
$N$ classes of particles (0,1,\ldots,N-1) with on-site hard-core exclusion.
  The number of particles of each specie is conserved. Consequently we can 
separate the Hilbert space into block disjoint sectors 
labeled by $(n_0,n_1,\ldots,n_{N-1})$, where $n_i=0,1,\ldots,L$ is the number 
of particle of specie i (i=0,1,...,N-1). The Hamiltonian has a $S_N$ 
symmetry due to its invariance under the permutation of particles 
species, that imply that all the energies can be obtained from the 
sectors $(n_0,n_1,\ldots,n_{N-1}) $, where $n_0 \le n_1 \le \cdots \le n_{N-1}$ and  
$n_0+n_1+\ldots+n_{N-1}=L$. Moreover the  quantum $SU(N)_q$ symmetry
of $H$ implies that all energies 
in the sector $(n_0^{\prime},n_1^{\prime},\ldots,n_{N-1}^{\prime})$ with 
$n_0^{\prime} \le n_1^{\prime} \le \cdots \le n_{N-1}^{\prime}$  are degenerated 
with the energies belonging to the sectors 
$(n_0,n_1,...,n_{N-1})$ with $n_0 \le n_1 \le \cdots \le n_{N-1}$, 
if $n_0^{\prime} \le n_0$ and
$n_0^{\prime} + n_1^{\prime} \le n_0 +n_1$ and so on up to $n_0^{\prime} +
n_1^{\prime}+\cdots+n_{N-2}^{\prime} \le n_0 +n_1+\cdots+n_{N-2}$.

 The nested Bethe ansatz equations (NBAE) for the 
$SU_q(N)$ Perk-Schultz model (\ref{H}) are given by 
(see e. g. ~\cite{resh,devega,freeBAE2,freeBAE})
\begin{eqnarray}
\label{ba3}
&&\prod_{j=1,j \ne i}^{p_k} F(u_i^{(k)},u_j^{(k)})= \prod_{j=1}^{p_{k-1}} 
f(u_i^{(k)},u_j^{(k-1)})  \prod_{j=1}^{p_{k+1}} f(u_i^{(k)},u_j^{(k+1)})  \end{eqnarray}
where $k=0,1,\ldots,N-2$ and $i=1,2,\ldots,p_k$.
The integer parameters $p_k$ depend on the particle occupation numbers
 $\{n_i\}$:
\begin{eqnarray}
\label{pk}
&&p_k=\sum_{i=0}^{k} n_i, \quad k=0,1,\ldots,N-2, \quad p_{-1}=0, 
\quad p_{N-1}=L,
\end{eqnarray}
and the functions $F(x,y)$ and  $f(x,y)$ are defined by
\begin{equation}
\label{ff}
F(x,y)=\frac{\cos(2y)-\cos(2x-2\eta)}{\cos(2y)-\cos(2x+2\eta)},\quad  \> f(x,y)=
\frac{\cos(2y)-\cos(2x-\eta)}{\cos(2y)-\cos(2x+\eta)}.
\end{equation}
In the NBAE  (\ref{ba3}) we have  variables of different classes.
The number of variables $u_{i}^{(k)}$ of class $k$ is equal to $p_k$ and the variables
$u^{(N-1)}_i=0\>(i=1,\ldots,L)$.
The whole system of NBAE consists of subsets of equations labelled by  $k$ 
and containing 
precisely $p_k$ equations $(k=0,1,\ldots,N-2)$.
 
The eigenenergies of the Hamiltonian (\ref{H}) in the sector $(n_0,n_1,\ldots,n_{N-1})$ are given
by 
\begin{equation}
\label{ba1}
E=-\sum_{j=1}^{p} \biggl(-q-\frac{1}{q}+\frac{\sin(u_j-\eta /2)}
{\sin(u_j+\eta /2)} +\frac{\sin(u_j+\eta /2)}
{\sin(u_j-\eta /2)} \biggr)
\end{equation}
where to simplify the notation $p \equiv p_{N-2}$ and $u_j \equiv u_j^{(N-2)}, j=1,2,\ldots,p$.

All the solutions of NBAE (\ref{ba3}) described in this and our previous paper  \cite{AS2}
satisfy the additional
"free-fermion" conditions (FFC)
\begin{eqnarray}
\label{ffc0}
&&f^L (u_i,0)=1 ,\quad i=1,\ldots,p.
\end{eqnarray}
 Consequently from (\ref{ba1}) and (\ref{ffc0}) the corresponding 
eigenenergies of the Hamiltonian (\ref{H}) are given by
\begin{equation}
\label{ba1f}
E=-2\sum_{j=1}^{p_{N-2}} 
\biggl(-\cos \eta + \cos \frac{\pi k_j}{L}\biggr),\>\>1\le k_j\le L-1
\end{equation}
where  $\{k_j\}$ is any set of distinct integers in the range 
 $1\le k_j\le L-1$. 

In the derivation of the  NBAE (\ref{ba3}) it is necessary to  find the 
eigenvalues of an  
 auxiliary  matrix that corresponds to the transfer matrix of an   
inhomogeneous $SU_q(N-1)$ invariant vertex model on the square 
lattice  of  width  $p$ in the horizontal direction. These eigenvalues enter into the NBAE
(see equation (51) from  \cite{freeBAE})\footnote{All details related to 
the construction of the transfer matrix can be found in reference 
\cite{freeBAE}.}:
\begin{eqnarray}
\label{Taux}
&&\Lambda_{\mbox{aux}}^{(N-1)}(u_i)=f^{-L}(u_i,0) \prod_{j=1,j \ne i}^p F(u_i,u_j), \quad
i=1,\ldots,p.
\end{eqnarray}
We see that the FFC (\ref{ffc0}) could be explained if there exists  an 
  special eigenvalue for this transfer matrix:
\begin{eqnarray}
\label{spec}
&&\Lambda_{\mbox{aux}}^{(N-1)}(u)=F^{-1}(u,u)\>\prod_{j=1}^p F(u,u_j).
\end{eqnarray} 
This observation can be formulated in the following conjecture.

{\it {\bf Conjecture:} let $q=\exp \frac{i\pi(N-1)}{N}$ and 
consider the inhomogeneous $SU_q(N-1)$ invariant vertex model 
on the  square lattice with p columns, where
$p=(N-1)\>k+r$, and $0 \le r \le N-2$.
The inhomogeneity of the model in the horizontal direction 
are fixed by the
vertical rapidities $u_j,\>j=1,\ldots,p$.
The  row-to-row transfer matrix, depending on the spectral 
parameter $u$ (horizontal rapidity), has a special factorizable eigenvalue given by formula
(\ref{spec}). The corresponding eigenvector belongs to the sector 
$S=\{n_0,n_1,\ldots,n_{N-2}\}$ 
where $n_i=k,\>i=0,\ldots,r-1$ and $n_i=k+1$ for $i=r,\ldots,N-2$. }

\vskip 1em

\begin{center}{\bf3. The $SU(3)$ Perk-Schultz model with periodic boundary}
\end{center}

The $SU(3)$ Perk-Schultz model \cite{PS} is the anisotropic version of the 
$SU(3)$ Sutherland model \cite{Sutherland}, with the Hamiltonian, in a periodic  L-site chain,
given by 
 \begin{eqnarray}
\label{H3}
&&H_{q} =\sum_{j=1}^{L} H_{j,j+1} 
\qquad ( H_{L,L+1} \equiv H_{L,1} ),  \\ 
&&H_{i,j} =-\sum_{a=0}^{1} \sum_{b=a+1}^{2} 
( E_i^{ab} E_{j}^{ba} + E_i^{ba} E_{j}^{ab}- q E_i^{aa} E_{j}^{bb} 
- 1/q  E_i^{bb} E_j^{aa} ). \nonumber
\end{eqnarray}

In our previous paper (see \cite{AS1} for details) it was shown for the 
periodic model  that:

{\it \noindent   The Hamiltonian (\ref{H3}) with L sites at $q=\exp(2 i \pi/3)$
has eigenvectors (not all of them) with energy and momentum given by 
\begin{eqnarray}
\label{C1}
&&E_I=-\sum_{j \in I} (1+2 \cos \frac{2 \pi j}{L}),  \\
\label{C2}
&& P_I=\frac{2 \pi}{L} \sum_{j \in I} j
\end{eqnarray}
with $I$ being a subset of $\cal I $ unequal elements of the 
set  $\{1,2,...,L\}$.
The number $\cal I$ has to be odd ${\cal I} =2k+1$ and 
the sector of appearance of the above levels  is  
$S_k \equiv (k,k+1,L-2k-1)$, $0 \le k \le (L-1)/2$.}

The corresponding  solutions 
of the NBAE were described in  \cite{AS2} and we intend now to present a procedure that allow us
to calculate 
the related wave functions. This procedure is based on the coordinate 
Bethe ansatz method and we follow here reference \cite{AB}. 

Due to the conservation of particles the total numbers of 
particles $n_0$,
$n_1$ and $n_2=L-n_0-n_1$ on classes 0,  1 and 2 
are good quantum numbers,  and consequently we can split the 
associated Hilbert space into block disjoint sectors labeled by the 
numbers $n_0$ and $n_1$.  We consider the eigenvalue equation

\begin{eqnarray}
\label{EE}
&& H|n_0,n_1>=E|n_0,n_1>
\end{eqnarray}
 where
\begin{eqnarray}
\label{EV}
&& |n_0,n_1>=\sum_{\{Q\}} 
\sum_{\{x\}} f(x_1,Q_1;\ldots;x_n,Q_n)|x_1,Q_1;\ldots;x_n,Q_n>
\end{eqnarray}
and $n = n_0 + n_1$. 
Here $|x_1,Q_1;\ldots;x_n,Q_n>$ means the configuration 
where a particle of class $Q_i$ ($Q_i=0,1$) is at position $x_i$ ($x_i=1,\ldots,L$). The summation
$\{Q\}=\{Q_1,\ldots,Q_n\}$ extends over all (0,1) sequences in which $n_0$ terms are 0 and $n_1$
terms are 1. The summation $\{x\}=\{x_1,\ldots,x_n\}$ runs, 
e $n$  increasing positive integers   with $1 \leq x_1 < \cdots<x_n\leq L$.
Before getting the results for general values of $n$ let us consider initially the simple cases 
where we have 1 or 2 particles.

{\it{\bf n=1.}}  For one particle on the chain (class 0 or 1), as a consequence
of the  translational invariance 
of (\ref{H3}) it is simple to verify directly that 
the eigenfunctions are the momentum-$k$ eigenfunctions 
\begin{eqnarray}
\label{n1a}
&&|1,0> = \sum_{x=1}^{L} f(x,0) |x,0> \quad \mbox {or} \quad 
|0,1> = \sum_{x=1}^{L} f(x,1) |x,1> 
\end{eqnarray}
with
\begin{eqnarray}
\label{n1b}
 &&f(x,0) =f(x,1) =  e^{ikx},\quad k= \frac{2\pi l}{L}, 
\quad l = 0,1,\ldots,L-1
\end{eqnarray}
and energy given by 
\begin{eqnarray}
\label{n1en}
&&E = e(k) \equiv  -e^{ik} - e^{-ik} + q + 1/q.
\end{eqnarray}

{\it {\bf  n =2.}} For two particles of classes $Q_1$ and 
$Q_2$ ($Q_1,Q_2=0,1$) on the lattice, the eigenvalue equation (\ref{EE}) 
 gives  us two distinct relations depending on the relative location of 
the particles. The first relation comes from the amplitudes where a 
particle of class $Q_1$ is at position $x_1$ and a 
particle $Q_2$ is at position $x_2$, where 
 $x_2 >x_1+1$. We obtain in this case the relation 
\begin{eqnarray}
\label{n2gen}
Ef(x_1,Q_1;x_2,Q_2) = &-&f(x_1-1,Q_1;x_2,Q_2) 
-f(x_1,Q_1;x_2+1,Q_2)  \nonumber \\ 
&-&f(x_1+1,Q_1;x_2,Q_2) 
 -f(x_1,Q_1;x_2-1,Q_2) \nonumber \\
&+&2\>(q+1/q)\>f(x_1,Q_1;x_2,Q_2).
\end{eqnarray}
This last equation can be solved promptly by the ansatz
\begin{eqnarray}
\label{n2pv}
f(x_1,Q_1;x_2,Q_2) = e^{ik_1x_1}e^{ik_2x_2}
\end{eqnarray}
with energy
\begin{eqnarray}
\label{n2en}
&&E = e(k_1) + e(k_2).
\end{eqnarray}
Since this relation  is symmetric under the interchange of  $k_1$ and $k_2$, 
 we can write 
a more general solution of (\ref{n2gen}) as
\begin{eqnarray}
\label{twopv}
f(x_1,Q_1;x_2,Q_2) &=& \sum_P A_{P_1,P_2}^{Q_1,Q_2}e^{i(k_{P_1}x_1 + 
k_{P_2}x_2)}   \nonumber \\
&=& A_{1,2}^{Q_1,Q_2}e^{i(k_1x_1+k_2x_2)} + 
A_{2,1}^{Q_1,Q_2}e^{i(k_2x_1+k_1x_2)}
\end{eqnarray}
with the same energy as in (\ref{n2en}). In the last equation the summation is over the
permutations  $P=P_1,P_2$ of (1,2). The second relation comes from the amplitude where  
$x_2 = x_1 + 1$ (matching condition). In this case instead of  (\ref{n2gen}) we have 
\begin{eqnarray}
\label{n2spec}
&&Ef(x_1,Q_1;x_1+1,Q_2) = -f(x_1-1,Q_1;x_1+1,Q_2) -f(x_1,Q_1;x_1+2,Q_2)
 \nonumber \\
&&-f(x_1,Q_2;x_1 +1,Q_1) + (2q+1/q) f(x_1,Q_1;x_1 +1,Q_2)
\quad Q_1 < Q_2 \nonumber \\
&&  \nonumber \\
&&Ef(x_1,Q_1;x_1+1,Q_2) = -f(x_1-1,Q_1;x_1+1,Q_2) -f(x_1,Q_1;x_1+2,Q_2)
 \nonumber \\
&&+ (q+1/q) f(x_1,Q_1;x_1 +1,Q_2)
\quad Q_1 = Q_2  \\
&&  \nonumber \\
&&Ef(x_1,Q_1;x_1+1,Q_2) = -f(x_1-1,Q_1;x_1+1,Q_2) -f(x_1,Q_1;x_1+2,Q_2)
 \nonumber \\
&&-f(x_1,Q_2;x_1 +1,Q_1) + (q+2/q) f(x_1,Q_1;x_1 +1,Q_2)
\quad Q_1 > Q_2. \nonumber
\end{eqnarray}
Since the ansatz (\ref{twopv}) with (\ref{n2en}) is also a solution of (\ref{n2gen}) with
$x_2=x_1+1$, we obtain from (\ref{n2spec}) 
\begin{eqnarray}
\label{n2dif}
&&f(x_1,Q_1;x_1,Q_2) + f(x_1+1,Q_1;x_1+1,Q_2)=
 \nonumber \\
&&1/q\>f(x_1,Q_1;x_1 +1,Q_2)+f(x_1,Q_2;x_1 +1,Q_1)
\quad Q_1 < Q_2 \nonumber \\
&&  \nonumber \\
&&f(x_1,Q_1;x_1,Q_2) + f(x_1+1,Q_1;x_1+1,Q_2)=
 \nonumber \\
&& (q+1/q) f(x_1,Q_1;x_1 +1,Q_2)
\quad Q_1 = Q_2  \\
&&  \nonumber \\
&&f(x_1,Q_1;x_1,Q_2) + f(x_1+1,Q_1;x_1+1,Q_2)=
 \nonumber \\
&&q\>f(x_1,Q_1;x_1 +1,Q_2)+f(x_1,Q_2;x_1 +1,Q_1) 
\quad Q_1 > Q_2. \nonumber
\end{eqnarray}
If we now substitute the ansatz (\ref{twopv}) into these equations the 
constants $A_{1,2}^{Q_1,Q_2}$ and $A_{2,1}^{Q_1,Q_2}$, initially arbitrary, should now satisfy 
\begin{eqnarray}
\label{n2A}
&&\sum_P\{ 
(\sigma_{P_1,P_2}+q 
e^{ik_{P_2}})\>A^{0,1}_{P_1,P_2}- e^{ik_{P_2}} A^{1,0}_{P_1,P_2}\}=0 \nonumber \\
&&\sum_P \sigma_{P_1,P_2}\>A^{Q,Q}_{P_1,P_2} =0 \quad Q=1,2\\
&&\sum_P\{ (\sigma_{P_1,P_2}+q^{-1} e^{ik_{P_2}})\>
A^{1,0}_{P_1,P_2}- e^{ik_{P_2}} A^{0,1}_{P_1,P_2}\}=0 \nonumber
\end{eqnarray}
where
\begin{eqnarray}
\label{sigma}
&&\sigma_{P_1,P_2}=1+e^{ik_{P_1}+ik_{P_2}}-(q+q^{-1})\>e^{ik_{P_2}}.
\end{eqnarray}
The system (\ref{n2A}) consists of 3 equations. The second equation can be easily rewritten as
\begin{eqnarray}
\label{S1}
&&A^{Q,Q}_{P_1,P_2} = -\frac{\sigma_{P_2,P_1}} 
{\sigma_{P_1,P_2}} A^{Q,Q}_{P_2,P_1}  
\end{eqnarray}
and combining the first and the third equations we obtain
\begin{eqnarray}
\label{S2}
A^{0,1}_{P_1,P_2} = &-&\frac{(1-q e^{ik_{P_2}})(1-q^{-1}e^{ik_{P_1}})} {\sigma_{P_1,P_2}}
A^{0,1}_{P_2,P_1} \nonumber \\  
&+&\frac{(e^{ik_{P_1}}-e^{ik_{P_2}})} {\sigma_{P_1,P_2}} A^{1,0}_{P_2,P_1} \\  
\label{S3}
A^{1,0}_{P_1,P_2} = &+&\frac{(e^{ik_{P_1}}-e^{ik_{P_2}})} {\sigma_{P_1,P_2}} A^{0,1}_{P_2,P_1}
\nonumber \\  
&-&\frac{(1-q e^{ik_{P_1}})(1-q^{-1}e^{ik_{P_2}})} {\sigma_{P_1,P_2}} A^{1,0}_{P_2,P_1}.  
\end{eqnarray}
The equations (\ref{S1}-\ref{S3}) can be written in a compact form
\begin{eqnarray}
\label{ASA}
A^{Q_1,Q_2}_{P_1,P_2} = - \sum_{Q_1^{\prime},Q_2^{\prime}=0}^1
S_{Q_2^{\prime},Q_1^{\prime}}^{Q_1,Q_2}(k_{P_1},k_{P_2}) 
A^{Q_1^{\prime},Q_2^{\prime}}_{P_2,P_1} \quad (Q_1,Q_2=0,1)  
\end{eqnarray}
where we have introduced the S-matrix
\begin{eqnarray}
\label{Smatrix}
S^{0,0}_{0,0}(k_1,k_2) &=&S^{1,1}_{1,1}(k_1,k_2)= 
\frac{\sigma_{2,1}} {\sigma_{1,2}}  \nonumber \\
S^{0,1}_{1,0}(k_1,k_2) &= &\frac{(1-q e^{ik_2})(1-q^{-1}e^{ik_1})} {\sigma_{1,2}}  \nonumber \\  
&& \\
S^{1,0}_{0,1}(k_1,k_2) & = &\frac{(1-q e^{ik_1})(1-q^{-1}e^{ik_2})} {\sigma_{1,2}}  \nonumber \\  
S^{0,1}_{0,1}(k_1,k_2) &=& S^{1,0}_{1,0}(k_1,k_2)= \frac{(e^{ik_2}-e^{ik_1})} {\sigma_{1,2}}.
\nonumber
\end{eqnarray}
The equations (\ref{S1}-\ref{S3}) or (\ref{ASA}) do not fix the "wave numbers" $k_1,k_2$.
In general, these numbers are fixed due to the cyclic boundary conditions:
\begin{eqnarray}
\label{pbc}
f(x_1,Q_1;x_2,Q_2)=f(x_2,Q_2;x_1+L,Q_1)
\end{eqnarray}
which from (\ref{twopv}) give us  the relations
\begin{eqnarray}
\label{Ashift}
A^{Q_1,Q_2}_{P_1,P_2} =  A^{Q_2,Q_1}_{P_2,P_1} e^{ik_{P_1}L}. 
\end{eqnarray}
This last equation, when solved by exploiting (\ref{ASA}) and (\ref{Smatrix}), gives us the 
possible values of $k_1$ and $k_2$, and from (\ref{n2en}) the eigenenergies 
in the sectors containing 2 particles. Instead of solving these equations for the 
particular case $n=2$ let us now consider the case of general $n$.

{\it {\bf General n.}} The above calculation can be generalized for arbitrary 
values of $n_0$ and $n_1$ of particles of classes 0 and 1, respectively 
($n_0 +n_1 = n$).  The ansatz for the wave function (\ref{EV})  becomes 
\begin{eqnarray}
\label{pwaves}
&&f(x_1,Q_1;\ldots;x_n,Q_n) = \sum_P A_{P_1,\ldots,P_n}^{Q_1,\cdots, Q_n}
 e^{i(k_{P_1}x_1+\cdots + 
k_{P_n}x_n)}
\end{eqnarray}
where the sum extends over all permutations $P$ of the integers 
$1,2,\ldots,n$. 
It is simple to see that the relations coming from the eigenvalue equation (\ref{EE}) for 
the components
 $|x_1,Q_1;\ldots;x_n,Q_n>$ where  
$x_{i+1} -x_i > 1$ for $i=1,2,\ldots,n$ are 
 satisfied by the ansatz (\ref{pwaves}) with energy 
\begin{eqnarray}
\label{nen}
&&E = \sum_{j=1}^n e(k_j).
\end{eqnarray}
On the other hand if
 a pair of particles belonging to classes  
$Q_i,Q_{i+1}$ is located at positions 
$x_i,\; x_{i+1}$, where $x_{i+1} = 
x_i + 1$, equation (\ref{EE}) with the ansatz (\ref{pwaves}) and the relation (\ref{nen}) give us
the generalization of relation (\ref{ASA}), namely
\begin{eqnarray}
\label{ASAgen}
A_{\ldots,P_i,P_{i+1},\ldots}^{\cdots, Q_i,Q_{i+1},\cdots} = 
 - \sum_{Q_1',Q_2'=0}^1 S_{Q_1',Q_2'}^{Q_i,Q_{i+1}}
(k_{P_i},k_{P_{i+1}}) A_{\ldots, P_{i+1},P_i,\ldots}^{\cdots, Q_2',Q_1',\cdots} 
\quad \;\; Q_i,Q_{i+1} =0,1
\end{eqnarray}
with $S$ given by (\ref{Smatrix}). Inserting the ansatz (\ref{pwaves}) in the boundary condition 
\begin{eqnarray}
\label{pbcgen}
f(x_1,Q_1;\ldots;x_n,Q_n) = f(x_2,Q_2;\ldots;x_n,Q_n;x_1+L,Q_1)
\end{eqnarray}
we obtain  the additional relation 
\begin{eqnarray}
\label{Ac}
A_{P_1,\ldots,P_n}^{Q_1,\cdots, Q_n} = e^{ik_{P_1}L}
A_{P_2,\ldots,P_n,P_1}^{Q_2,\cdots, Q_n,Q_1}
\end{eqnarray}
which together with (\ref{ASAgen}) should give us the eigenenergies.

 Successive applications of (\ref{ASAgen}) give us in 
general distinct relations between the amplitudes. 
They are consistent because as we will see 
below the $S$ matrix (\ref{Smatrix}) coincides with  the famous 6-vertex $R$ matrix and  satisfies
the Yang-Baxter equation. Hence we may use these relations to obtain the eigenenergies of the
Hamiltonian (\ref{H3}). Applying the relation  
(\ref{ASAgen}) $n$ times  we obtain from (\ref{Ac}) a relation 
connecting the amplitudes with the same momenta, namely,
\begin{eqnarray} 
\label{ATA}
&&A_{P_1,\ldots,P_n}^{Q_1,\ldots,Q_n} = e^{ik_{P_1}L} A_{P_2,\ldots,P_n,P_1}^
{Q_2,\ldots,Q_n,Q_1} = (-1)^{n-1}  
e^{ik_{P_1}L} \nonumber \\
&&\sum_{Q_1',\ldots,Q_n'}\sum_{Q_1'',\ldots,Q_n''} 
S_{Q_1',Q_1''}^{Q_1,Q_2''}(k_{P_1},k_{P_1})
S_{Q_2',Q_2''}^{Q_2,Q_3''}(k_{P_2},k_{P_1})\cdots \\
&&S_{Q_{n-1}',Q_{n-1}''}^{Q_{n-1},Q_n''}(k_{P_{n-1}},k_{P_1})
S_{Q_{n}',Q_{n}''}^{Q_{n},Q_1''}(k_{P_{n}},k_{P_1})
A_{P_1,\ldots,P_n}^{Q_1',\ldots,Q_n'}\nonumber
\end{eqnarray}
where we have introduced the harmless extra sum
\begin{eqnarray}
\label{extra}
1 = \sum_{Q_1'',Q_2''=0}^1 \delta_{Q_2'',Q_1'} \delta_{Q_1'',Q_1} =
 \sum_{Q_1'',Q_2''=0}^1 S_{Q_1',Q_1''}^{Q_1,Q_2''}(k_{P_1},k_{P_1}).
\end{eqnarray}
In order to fix the values of $\{k_j\}$ we should solve (\ref{ATA}), i.e., 
we should find the eigenvalues $\Lambda(k)$ of the matrix
\begin{eqnarray}
\label{Tmatrix}
{\cal T}(k)_{\{Q'\}}^{\{Q\}} = \sum_{Q_1'',\ldots,Q_n''=0}^1 \left\{ \left(
\prod_{l=1}^{n-1} 
S_{Q_l',Q_l''}^{Q_l,Q_{l+1}''}(k_{P_l} ,k)\right) 
S_{Q_n',Q_n''}^{Q_n,Q_1''}(k_{P_n},k) \right\}
\end{eqnarray}
and the Bethe-ansatz equations which fix the set $\{k_l\}$ will be given from
 (\ref{ATA}) by 
\begin{eqnarray}
 \label{aux}
e^{-ik_jL} = (-1)^{n-1} \Lambda(k_j) \quad j=1,\ldots,n.
\end{eqnarray}
We identify ${\cal T}(k)$ as the transfer matrix of an inhomogeneous 
6-vertex model, on a periodic lattice,
with Boltzmann weights  $S_{Q_1',Q_2'}^{Q_1,Q_2}(k_{P_l},k)$ ($l=1,\ldots,n$).  
Consequently, in order to obtain the eigenenergies of the quantum 
chain (\ref{H3}) we should diagonalize the above transfer matrix $T(k)$.

It is convenient to introduce the variables $\{u_j\}$ as in the NBAE of 
\S 2:
\begin{eqnarray}
 \label{u}
e^{-ik} = \frac{\sin(u-\pi/3)}{\sin(u+\pi/3)}.
\end{eqnarray}
In terms of these variables the  S-matrix  (\ref{Smatrix}) have a  
difference form
\begin{eqnarray}
\label{Smatrixu}
S^{0,0}_{0,0}(u_1-u_2) &=&S^{1,1}_{1,1}(u_1-u_2)= 
\frac{\sin (u_1-u_2+\pi/3)} {\sin (u_2-u_1+\pi/3)},  \nonumber \\
S^{0,1}_{1,0}(u_1-u_2) &= &\frac{\sqrt{3} e^{i(u_1-u_2)}} {2 \sin (u_2-u_1+\pi/3)} \nonumber \\  
&& \\
S^{1,0}_{0,1}(u_1-u_2) &= &\frac{\sqrt{3} e^{i(u_2-u_1)}} {2 \sin (u_2-u_1+\pi/3)} \nonumber \\  
S^{0,1}_{0,1}(u_1-u_2) &=& S^{1,0}_{1,0}(u_1-u_2)=\frac{\sin (u_2-u_1)} {\sin (u_2-u_1+\pi/3)} .
\nonumber
\end{eqnarray}
where we should remind that we are considering $q=\exp(2i\pi/3)$.

Introducing for convenience the new amplitudes 
\begin{eqnarray}
\label{Anew}
A_{P_1,\ldots,P_n}^{Q_1,\ldots, Q_n} = \exp(i\sum_{j=1}^n  \delta_{Q_j,0}\> u_{P_j})
\tilde A_{P_1,\ldots,P_n}^{Q_1,\ldots, Q_n}
\end{eqnarray}
 we obtain instead of (\ref{ASAgen}) a similar relation  
with $\tilde A$ instead of $A$ and 
with the $\tilde S$-matrix of the symmetric 6-vertex model, 
whose non-zero components are given by  
 \begin{eqnarray}
\label{Smatrixsym}
\tilde S^{0,0}_{0,0}(u) &=&\tilde S^{1,1}_{1,1}(u)= 
\rho \> \sin (\pi/3-u)   \nonumber \\
\tilde S^{0,1}_{0,1}(u) &=& \tilde S^{1,0}_{1,0}(u)=\rho \> \sin u \\
\tilde S^{0,1}_{1,0}(u) &= &\tilde S^{1,0}_{0,1}(u)=\rho \>  \sin \pi/3  \nonumber 
\end{eqnarray}
where $\rho = 1/ \sin (\pi/3+u)$.

The  matrix (\ref{Tmatrix}) is the transfer matrix of an inhomogeneous  6-vertex model  on the
square lattice of  width $n=n_0+n_1$.
In the next section we show that for the case where $n_0=k$ and $n_1=k+1$
(sector (k,k+1) of the 6-vertex model on the lattice of width $2k+1$) 
this transfer matrix has the special 
eigenvalue $\Lambda=1$ 
independently on the values of the 
parameters $u$ and $u_j$ ($j=1,\ldots,2n+1$).
Consequently,  since $n=2k+1$,   (\ref{aux}) reduces to
\begin{eqnarray}
 \label{FFC}
e^{-ik_jL} = 1, \quad j=1,\ldots,n,
\end{eqnarray}
and the associated energies of the quantum chain (\ref{H3}) are free-fermion-like.

\vspace{0.5cm}

\begin{center}{\bf4. The special eigenvalue}
\end{center}

Applying the Bethe ansatz  method to the 
auxiliary transfer matrix ${\cal T}(k)={\cal T}(u)$  
introduced  in (\ref{Tmatrix}) 
 one obtain the well known NBAE 
\cite{resh,devega}, which were considered in our 
paper \cite{AS2}. We have shown there that these 
NBAE are consistent with the FFC (\ref{FFC}) for the 
sectors $(n_0,n_1,n_2)=(k,k+1,L-2k-1)$.
However it  is more convenient 
here to follow an early paper of Baxter \cite{B} who 
considered the most general integrable inhomogeneous 
6-vertex model. Any  eigenvalue $T(u)$ of the inhomogeneous 
model with $\rho=1$ in (\ref{Smatrixsym})
satisfy the  equation
 \begin{eqnarray}
\label{TQ}
T(u)\>Q(u) =\biggl (\prod_{j=1}^n \sin (\pi/3-u+u_j)\biggr )\>Q(u-2\pi/3) \nonumber \\
+ \biggl ( \prod_{j=1}^n \sin (u-u_j) \biggr )\>Q(u+2\pi/3) 
\end{eqnarray}
where  $Q(u)$ is an 
auxiliary trigonometric polynomial of  degree $n_0$, namely,
 \begin{eqnarray}
\label{Q}
Q(u)=\prod_{j=1}^{n_0} \sin (u-\alpha_j).
\end{eqnarray}
It is clear from (\ref{TQ}) that 
 $T(u)$ is a  trigonometric polynomial of  degree $n$ and
from (\ref{Smatrixsym})
the   eigenvalues of (\ref{Tmatrix}) (where now $\rho \neq 1$) are given by
  \begin{eqnarray}
\label{LambdaT}
\Lambda(u)=T(u)/\prod_{j=1}^n \sin (\pi/3+u-u_j).
\end{eqnarray}
This last expression implies that we have an eigenvalue $\Lambda(u) =1$ if 
 \begin{eqnarray}
\label{Tour}
T(u)=\prod_{j=1}^n \sin (\pi/3+u-u_j).
\end{eqnarray}
The existence of this special eigenvalue was argued by Baxter for the 
more general case of the 8-vertex model with 
special values of the crossing parameters \cite{Bold}. 
We can prove the existence of this eigenvalue for 
lattices with odd values of its width $n$. 
In this case  we can rewrite Baxter's T-Q equation (\ref{TQ})  as 
 \begin{eqnarray}
\label{cycle}
f(u)+f(u+2\pi/3) + f(u-2\pi/3)=0
\end{eqnarray}
where 
 \begin{eqnarray}
\label{f}
f(u) \equiv Q(u+2\pi/3)\>\prod_{j=1}^n \sin (u-u_j).
\end{eqnarray}
It is now clear  that $f(u)$ is a trigonometric polynomial of the degree $n+n_0$.
Equation (\ref{f}) coincides with  (6) in \cite{Snew}, where
$f(u)$ is also a trigonometric polynomial. 

In \cite{Snew} it is shown that 
for any set of complex numbers $u_j$, $j=1,\ldots,2k+1$, there exists 
a trigonometric polynomial $Z(u)$ of degree $n$ on the  variable $u$ 
 such
that 
 \begin{eqnarray}
\label{fff}
f(u) \equiv Z(u)\>\prod_{j=1}^{2k+1} \sin (u-u_j)
\end{eqnarray}
satisfies (\ref{cycle})\footnote{$Z(u)$ is the partition 
function of the inhomogeneous 6-vertex 
model with domain wall boundary conditions and with rapidities 
$\{u,u_j\},\quad j=1,\ldots,2k+1$.}.
The degree of $f$ is equal to $n+n_0=3k+1$ and the degree of $Z$ is 
equal to the degree of $Q$, so that $k=n_0$. We obtain consequently 
 that the construction of paper 
\cite{Snew} correspond to $n_0=k$ and $n=n_0+n_1=2k+1$, i.e. $n_1=k+1$, 
that complete the proof of existence of the special eigenvalue 
(\ref{Tour}) for odd 
values of $n$. 

\vspace{0.5cm}

\begin{center}{\bf5. The special wave function of the inhomogeneous 6-vertex
model at $q^{2i\pi/3}$}
\end{center}

We consider several families of transfer matrices (\ref{Tmatrix}) corresponding to each distinct
permutation 
$P=\{P_1,\ldots,P_n\}$.  It follows from  (\ref{ATA}) that 
$A_{P_1,\ldots,P_n}^{Q_1,\cdots, Q_n}$ are the $2^n$ components ($ Q_1,...,Q_n=0,1$) of an
eigenvector of the transfer matrix (\ref{Tmatrix}). Now we are going to investigate these
components for our special eigenvalue $\Lambda=1$ using the  generalization
arbitrary number of particles $n$, i. e.,
\begin{eqnarray}
\label{nAgen}
&&\sum_{P=\{P_l,P_{l+1}\}}\{ (\sigma_{P_l,P_{l+1}}+q
e^{ik_{P_{l+1}}})\>A^{\cdots,0,1,\cdots}_{\ldots,P_l,P_{l+1},\ldots}- e^{ik_{P_{l+1}}}
A^{\cdots,1,0,\cdots}_{\ldots,P_l,P_{l+1},\ldots}\}=0 \nonumber \\
&&\sum_{P=\{P_l,P_{l+1}\}} \sigma_{P_l,P_{l+1}}\>A^{\cdots,Q,Q,\cdots}_{\ldots,P_l,P_{l+1},\ldots}
=0\quad Q=1,2 \\
&&\sum_{P=\{P_l,P_{l+1}\}}\{ (\sigma_{P_l,P_{l+1}}+q^{-1} 
e^{ik_{P_{l+1}}})\>A^{\cdots,1,0,\cdots}_{\ldots,P_l,P_{l+1},\ldots}- 
e^{ik_{P_{l+1}}} 
A^{\cdots,0,1,\cdots}_{\ldots,P_l,P_{l+1},\ldots}\}=0 \nonumber
\end{eqnarray}
where all indices shown by dots are fixed.
When we constructed the $S$-matrix in (\ref{Smatrix}) we expressed $A_{P_1,P_2}^{Q_1,Q_2}$ 
as a linear 
combination of  $A_{P_2,P_1}^{Q_1,Q_2}$ and  $A_{P_2,P_1}^{Q_2,Q_1}$. 
Now, on the other hand, we intend to express $A_{P_1,P_2}^{Q_1,Q_2}$ as a linear combination of
$A_{P_1,P_2}^{Q_2,Q_1}$ and  $A_{P_2,P_1}^{Q_2,Q_1}$  ($Q_1 \ne Q_2$). 
Combining the first and the third equation of the set  (\ref{n2A}) we obtain
\begin{eqnarray}
\label{SS1}
A^{1,0}_{P_1,P_2} ( e^{ik_{P_2}}-e^{ik_{P_1}}) &=&(1-q e^{ik_{P_1}})(1-q^{-1}e^{ik_{P_2}})
A^{0,1}_{P_1,P_2} \nonumber \\  
&+&\sigma_{P_2,P_1} A^{0,1}_{P_2,P_1} \\  
A^{0,1}_{P_1,P_2} ( e^{ik_{P_2}}-e^{ik_{P_1}}) &=&(1-q^{-1} e^{ik_{P_1}})(1-q e^{ik_{P_2}})
A^{1,0}_{P_1,P_2} \nonumber \\  
&+&\sigma_{P_2,P_1} A^{1,0}_{P_2,P_1}.   \nonumber
\end{eqnarray}
Changing the variables as in (\ref{u}) these equations are replaced by 
\begin{eqnarray}
\label{Snew1}
A^{1,0}_{P_1,P_2} \sin(u_{P_2}-u_{P_1}) &=& \frac{\sqrt{3}}{2} e^{i (u_{P_2}-u_{P_1})}
A^{0,1}_{P_1,P_2} \nonumber \\  
&+&\sin (u_{P_1}-u_{P_2} + \pi/3) A^{0,1}_{P_2,P_1} \\  
A^{0,1}_{P_1,P_2} \sin(u_{P_2}-u_{P_1}) &=& \frac{\sqrt{3}}{2} e^{i (u_{P_1}-u_{P_2})}
A^{1,0}_{P_1,P_2} \nonumber \\  
&+&\sin(u_{P_1}-u_{P_2} + \pi/3) A^{1,0}_{P_2,P_1}.\nonumber  
\end{eqnarray}
Generalizing these equations for  arbitrary $n$ and considering $\tilde A$ instead of $A$ (as in
(\ref{Anew})) we obtain
\begin{eqnarray}
\label{work1}
\tilde A^{\cdots,1,0,\cdots}_{\ldots,P_l,P_{l+1},\ldots} \sin(u_{P_{l+1}}-u_{P_l})&=&
 \frac{\sqrt{3}}{2} \tilde A^{\cdots,0,1,\cdots}_{\ldots,P_l,P_{l+1}\ldots} \nonumber \\
&+&\sin(u_{P_l}-u_{P_{l+1}}+\pi/3) 
\tilde A^{\cdots,0,1,\cdots}_{\ldots,P_{l+1},P_l,\ldots} =0 \nonumber \\
\tilde A^{\cdots,0,1,\cdots}_
{\ldots,P_l,P_{l+1},\ldots} \sin(u_{P_{l+1}}-u_{P_l})&=&
 \frac{\sqrt{3}}{2} \tilde A^{\cdots,1,0,\cdots}_{\ldots,P_l,P_{l+1},\ldots}  \\
&+&\sin(u_{P_l}-u_{P_{l+1}}+\pi/3) \tilde A^{\cdots,1,0,\cdots}_{\ldots,P_{l+1},P_l,\ldots} =0.
\nonumber 
\end{eqnarray}
We have also a generalization of the second equation in  (\ref{n2A})

\begin{eqnarray}
\label{work2}
&& \sin(u_{P_{l+1}}-u_{P_l}+\pi/3) \>\tilde A^{\cdots,Q,Q,\cdots}_{\ldots,P_l,P_{l+1},\ldots} =
\nonumber \\ &&-\sin(u_{P_l}-u_{P_{l+1}}+\pi/3) \>\tilde
A^{\cdots,Q,Q,\cdots}_{\ldots,P_{l+1},P_l,\ldots} \quad (Q=0,1).
\end{eqnarray}
Moreover due to FFC (\ref{FFC}) equation (\ref{Ac}) leads to the cyclic symmetry 
\begin{eqnarray}
\label{Acycle}
A_{P_1,\ldots,P_n}^{Q_1,\cdots, Q_n} = 
A_{P_2,\ldots,P_n,P_1}^{Q_2,\cdots, Q_n,Q_1}
\end{eqnarray}
which is also valid for $\tilde A$.

Let us begin with the simplest nontrivial case $k=1$, where the sector of appearance of
free-fermion levels is $(1,2,L-3)$.
Equation (\ref{EV}) for the eigenvectors can be written as follows 
\begin{eqnarray}
\label{EV3}
&& |n_0,n_1>= |1,2> = \nonumber \\
&& \sum_{\{Q\}} \sum_{1 \le x_1 ,x_2 ,x_3 \le L} 
f(x_1,Q_1;x_2,Q_2;x_3,Q_3)|x_1,Q_1;x_2,Q_2;x_3,Q_3>
\end{eqnarray}
where  we  sum over the three sequences of $\{Q\}$:
$\{0,1,1\}$, $\{1,0,1\}$ and $\{1,1,0\}$ that  correspond to 
the configurations where $n_0=1$, $n_1=2$
$(n=n_0+n_1=3)$.
The amplitudes are given by the ansatz (\ref{pwaves}) 
\begin{eqnarray}
\label{pwaves3}
&&f(x_1,Q_1;x_2,Q_2;x_3,Q_3) = \sum_P A_{P_1,P_2,P_3}^{Q_1,Q_2,Q_3}
e^{i(k_{P_1}x_1+k_{P_2}x_2+k_{P_3}x_3)}.
\end{eqnarray}
In the above expression there are 3 types of  parameters  $\{A\}$ 
which are related among themselves by the cyclic symmetry:
\begin{eqnarray}
\label{Acycle3}
\tilde A_{P_1,P_2,P_3}^{0,1,1} = \tilde A_{P_2,P_3,P_1}^{1,1,0} = \tilde A_{P_3,P_1,P_2}^{1,0,1}.
\end{eqnarray}
Before proceding let us introduce the simplified notations 
\begin{eqnarray}
\label{ShortN}
&&s_{P_1,P_2}= 
\frac{\sin (u_{P_1}-u_{P_2}+\pi/3)}{\sin (\pi/3)} \nonumber \\
&&d_{P_1,P_2}= \frac{\sin (u_{P_1}-u_{P_2})}{\sin (\pi/3)}.
\end{eqnarray}
The relation  (\ref{work2}) can be reduced to the equation
\begin{eqnarray}
\label{work23}
 s_{P_3,P_2}\>\tilde A^{0,1,1}_{P_1,P_2,P_3} 
=- s_{P_2,P_3}\>\tilde A^{0,1,1}_{P_1,P_3,P_2}
\end{eqnarray}
so that due to (\ref{Acycle3}) 
\begin{eqnarray}
\label{A1}
\tilde A^{0,1,1}_{P_1,P_2,P_3} = \pm C\{P_1\} s_{P_2,P_3}
\end{eqnarray}
where the sign depends on the parity of the permutation $P=\{P_1,P_2,P_3\}$
and $C\{i\}, i=1,2,3$ are  unknown coefficients.

The relations (\ref{work1}) give us in particular the equation
\begin{eqnarray}
\label{work13}
\tilde A^{1,0,1}_{P_1,P_2,P_3} d_{P_2,P_1}=
\tilde A^{0,1,1}_{P_1,P_2,P_3}+ s_{P_1,P_2} \tilde A^{0,1,1}_{P_2,P_1,P_3}. \end{eqnarray}
Using the cyclic symmetry (\ref{Acycle3}) and (\ref{A1}) we obtain
\begin{eqnarray}
\label{eq1}
C\{P_2\} s_{P_3,P_1} d_{P_2,P_1}= C\{P_1\} s_{P_2,P_3} - C\{P_2\} s_{P_1,P_2} s_{P_1,P_3} 
 \end{eqnarray}
that together with the identity
\begin{eqnarray}
\label{id1}
s_{P_3,P_1} d_{P_2,P_1} +  s_{P_1,P_2} s_{P_1,P_3} = s_{P_2,P_3}
 \end{eqnarray}
give us  $C\{P_2\}=C\{P_2\}=C$, i.e., up to a normalization factor we have
\begin{eqnarray}
\label{res1}
\tilde A_{P_1,P_2,P_3}^{0,1,1} = \pm s_{P_2,P_3}
\end{eqnarray}
where the sign depends on the parity of the permutation $P$.

Consider further the next sector 
$(2,3,L-5)$, i.e., $n_0=2$, $n_1=3$ and $n=n_0+n_1=5$.  
In this case we have two sets of $\{\tilde A\}$, which are 
related due to the cyclic symmetry, namely,  

\begin{eqnarray}
\label{cycle5}
&& \tilde A^{0,0,1,1,1}_{P_1,P_2,P_3,P_4,P_5}=
\tilde A^{0,1,1,1,0}_{P_2,P_3,P_4,P_5,P_1}= \dots =
\tilde A^{1,0,0,1,1}_{P_5,P_1,P_2,P_3,P_4} \nonumber \\
&& \tilde A^{0,1,0,1,1}_{P_1,P_2,P_3,P_4,P_5}=
\tilde A^{1,0,1,1,0}_{P_2,P_3,P_4,P_5,P_1}= \dots =
\tilde A^{1,0,1,0,1}_{P_5,P_1,P_2,P_3,P_4}.
\end{eqnarray}
Let us begin with the first set. The relation (\ref{work2}) 
is solved by the ansatz 
\begin{eqnarray}
\label{A2}
 \tilde 
A^{0,0,1,1,1}_{P_1,P_2,P_3,P_4,P_5} = \pm 
C\{P_1,P_2\} s_{P_1,P_2}
 s_{P_3,P_4} s_{P_3,P_5} s_{P_4,P_5} 
\end{eqnarray}
where the sign depends on the parity of the permutation $P=\{P_1,P_2,P_3,P_4,P_5\}$
and $C\{i,j\}, i,j=1,2,3,4,5$ are symmetric unknown coefficients $C\{i,j\}=C\{j,i\}$.
The system (\ref{work1}) contains in particular the three following equations:
\begin{eqnarray}
\label{inisys}
&&\tilde A^{0,1,0,1,1}_{P_1,P_2,P_3,P_4,P_5} d_{P_3,P_2} = 
\tilde A^{0,0,1,1,1}_{P_1,P_2,P_3,P_4,P_5} + 
s_{P_2,P_3} A^{0,0,1,1,1}_{P_1,P_3,P_2,P_4,P_5} \nonumber \\
&&\tilde A^{0,1,0,1,1}_{P_2,P_1,P_3,P_4,P_5} d_{P_3,P_1} = 
\tilde A^{0,0,1,1,1}_{P_2,P_1,P_3,P_4,P_5} + 
s_{P_1,P_3} A^{0,0,1,1,1}_{P_2,P_3,P_1,P_4,P_5}  \\
&&\tilde A^{1,0,0,1,1}_{P_1,P_2,P_3,P_4,P_5} d_{P_2,P_1} = 
\tilde A^{0,1,0,1,1}_{P_1,P_2,P_3,P_4,P_5} + 
s_{P_1,P_2} \tilde A^{0,1,0,1,1}_{P_2,P_1,P_3,P_4,P_5}. \nonumber 
\end{eqnarray} 
Excluding from this system $\tilde A^{0,1,0,1,1}_{\dots}$,
 using the cyclic symmetry 

$$\tilde A^{1,0,0,1,1}_{P_1,P_2,P_3,P_4,P_5} 
=\tilde A^{0,0,1,1,1}_{P_2,P_3,P_4,P_5,P_1}$$

\noindent and limiting ourselves with the unit permutation 
 we get
\begin{eqnarray}
\label{main5}
&&\tilde A^{0,0,1,1,1}_{2,3,4,5,1} d_{2,1} d_{3,2} d_{3,1} = 
(\tilde A^{0,0,1,1,1}_{1,2,3,4,5} + s_{2,3} 
\tilde A^{0,0,1,1,1}_{1,3,2,4,5}) d_{3,1} + \nonumber \\
&&( \tilde A^{0,0,1,1,1}_{2,1,3,4,5} + s_{1,3} 
\tilde A^{0,0,1,1,1}_{2,3,1,4,5}) d_{3,2} s_{1,2}. 
\end{eqnarray}
Inserting here the ansatz (\ref{A2}) and using the antisymmetry of $d_{i,j}$ we obtain 
\begin{eqnarray}
\label{aux1}
&& C\{1,2\} s_{1,2} s_{3,4} s_{3,5} (d_{2,3} s_{2,1} - d_{1,3}) +
 C\{1,3\} d_{1,3} s_{1,3} s_{2,3} s_{2,4} s_{2,5} + \nonumber \\
&&C\{2,3\} d_{2,3} s_{2,3} (d_{1,2} d_{1,3} s_{4,1} s_{5,1} - s_{1,2} s_{1,3} s_{1,4} s_{1,5})=0.
\end{eqnarray}
Using the identity
\begin{eqnarray}
\label{id2}
d_{2,3} s_{2,1} -  d_{1,3} = - d_{1,2} s_{2,3}
 \end{eqnarray}
and removing a common multiplier $s_{2,3}$ from (\ref{aux1}) we obtain the more simple equation
\begin{eqnarray}
\label{aux2}
&& -C\{1,2\} d_{1,2} s_{1,2} s_{3,4} s_{3,5} +
 C\{1,3\} d_{1,3} s_{1,3} s_{2,4} s_{2,5} + \nonumber \\
&&C\{2,3\} d_{2,3} (d_{1,2} d_{1,3} s_{4,1} s_{5,1} - s_{1,2} s_{1,3} s_{1,4} s_{1,5})=0.
\end{eqnarray}
By interchaging indices 1 and 2 we obtain a distinct equation. 
If we now exclude $C\{1,2\}$ from these two equations we get a 
relation between   
$C\{2,3\}$ and $C\{1,3\}$, namely,
\begin{eqnarray}
\label{aux3}
&& C\{2,3\} d_{2,3} (d_{1,2} d_{1,3} s_{2,1} s_{4,1} s_{5,1} +
  s_{1,2} s_{1,4} s_{1,5} (s_{2,3} - s_{2,1} s_{1,3})) = \nonumber \\
&& C\{1,3\} d_{1,3} (d_{1,2} d_{2,3} s_{1,2} s_{4,2} s_{5,2} +
  s_{2,1} s_{2,4} s_{2,5} (s_{1,2} s_{2,3} - s_{1,3})).
\end{eqnarray}
This last relation takes a nice form if we use the identities 
\begin{eqnarray}
\label{id3}
s_{2,3} - s_{2,1} s_{1,3} = d_{1,2} d_{1,3} \quad \quad 
s_{1,2} s_{2,3} - s_{1,3} = d_{1,2} d_{2,3} 
 \end{eqnarray} 
and remove the  common factors $d_{1,2}$, $d_{1,3}$ and
$d_{2,3}$, i. e.,   
\begin{eqnarray}
\label{aux4}
&& C\{2,3\} ( s_{2,1} s_{4,1} s_{5,1} +
  s_{1,2} s_{1,4} s_{1,5}) = \nonumber \\
&& C\{1,3\} ( s_{1,2} s_{4,2} s_{5,2} +
  s_{2,1} s_{2,4} s_{2,5}).
\end{eqnarray}
Using standard trigonometric identities one can show that the left side combination
\begin{eqnarray}
\label{id4}
&& s_{2,1} s_{4,1} s_{5,1} +
  s_{1,2} s_{1,4} s_{1,5} = \frac{2}{3} \{\cos(u_1+u_2-u_4-u_5)+ \nonumber \\
&&\cos(u_1-u_2+u_4-u_5) + \cos(u_1-u_2-u_4+u_5)\}
\end{eqnarray}
 has a $S_4$ symmetry, and consequently   (\ref{aux4}) reduces to 
\begin{eqnarray}
\label{aux5}
&& C\{2,3\}  = C\{1,3\}.
\end{eqnarray}
This means that $C$ does not depend on its indices and we have up to a normalization factor 
\begin{eqnarray}
\label{A5}
 \tilde A^{0,0,1,1,1}_{P_1,P_2,P_3,P_4,P_5} = \pm s_{P_1,P_2}
 s_{P_3,P_4} s_{P_3,P_5} s_{P_4,P_5}. 
\end{eqnarray}

The second set of amplitudes can be found from the first relation
in (\ref{inisys}). For example
\begin{eqnarray}
\label{aux6}
 \tilde A^{0,1,0,1,1}_{1,2,3,4,5} = d_{2,3}^{-1} s_{4,5} (s_{1,3}
 s_{2,3} s_{2,4} s_{2,5} - s_{1,2} s_{3,4} s_{3,5}) \equiv \nonumber \\
 d_{2,3}^{-1} s_{4,5} \{s_{2,3} s_{2,5} (s_{1,3} s_{2,4} - s_{1,2} s_{3,4})+
 s_{1,2} s_{3,4} (s_{2,3} s_{2,5} - s_{3,5})\}
\end{eqnarray}
and using the identities
\begin{eqnarray}
\label{id5}
s_{1,3} s_{2,4} - s_{1,2} s_{3,4} = d_{2,3} s_{4,1} \quad
 \quad  s_{2,3} s_{2,5} - s_{3,5} =d_{2,3} s_{5,2}
\end{eqnarray}
we obtain 
\begin{eqnarray}
\label{A51}
 \tilde A^{0,1,0,1,1}_{1,2,3,4,5} = s_{4,5} (s_{2,3}
 s_{2,5} s_{4,1} + s_{1,2} s_{3,4} s_{5,2}), 
\end{eqnarray}
or equivalently by using some additional identities we  get  
\begin{eqnarray}
\label{A52}
 \tilde A^{0,1,0,1,1}_{1,2,3,4,5} = s_{4,5} (s_{2,3}
 s_{2,4} s_{5,1} + s_{1,2} s_{3,5} s_{4,2}). 
\end{eqnarray}
In a similar way we can derive  the general answer
\begin{eqnarray}
\label{A5gen}
&& \tilde A^{0,1,0,1,1}_{P_1,P_2,P_3,P_4,P_5}/s_{P_4,P_5} = 
 s_{P_2,P_3} s_{P_2,P_5} s_{P_4,P_1} + s_{P_1,P_2} s_{P_3,P_4} s_{P_5,P_2}= \nonumber \\
&& s_{P_2,P_3} s_{P_2,P_4} s_{P_5,P_1} + 
s_{P_1,P_2} s_{P_3,P_5} s_{P_4,P_2}.
\end{eqnarray}

Equations (\ref{res1}) and (\ref{A5}) induce us to conjecture
that for an arbitrary $n$ the amplitudes $\{\tilde A\}$ 
of the special wave function 
of the inhomogeneous 6-vertex model are given by the ansatz 
\begin{eqnarray}
\label{Angen}
&& A^{0,\dots,0,1,\dots,1}_{P_1,\dots,P_k,P_{k+1},\dots,P_{2k+1}}=
\prod_{1 \le i < j \le k} s_{P_i,P_j} \> \prod_{k+1 \le i < j \le 2k+1}  s_{P_i,P_j}
\end{eqnarray}
where $k$ and $k+1$ are the  numbers of particles of species  0 and 1, 
respectively.
We checked this formula for $n=7$ analytically and for $n=9$ using a bruteforce diagonalization.
It is a challenge to prove the validity of this formula for an arbitrary odd number $n$.
The remaining amplitudes  can be found  by using this ansatz and 
sucessive application of the "recursion" relation (\ref{work1}).
This completes our discussion of the $SU(2)$ periodic case.

\vspace{0.5cm}

\begin{center}{\bf 6. Summary and Conclusions} 
\end{center}

In previous papers \cite{AS1,AS2} it was shown the existence of 
free-fermion-like 
 energies for the anisotropic $SU(N)$ Perk-Schultz model with anisotropy
parameter $q=\exp{i\pi(N-1)/N}$. These solutions were found for general values
of $N$ in the case of free boundary condition, where the model is $SU_q(N)$
invariant and for the $SU(3)$ case in the periodic case.

In \S 2 of this paper we show that the above observations, for the case of free
boundaries, could be explained by a conjecture stating the existence of a
special factorizable eigenvalue of the auxiliary inhomogeneous transfer matrix
of a $SU_q(N-1)$ vertex model with the same value of the anisotropy. Although
we believe that a general derivation of such factorizable eigenvalue would be
possible we restricted our analytical work in the simplest case of the periodic
$SU(3)$ Perk-Schultz model at $q=\exp{i\frac{2\pi}{3}}$. In this case the
associated transfer matrix is that of the inhomogeneous 6-vertex model and 
 the existence of the factorizable eigenvalue, as shown in \S 4, follows from the
T-Q Baxter equation (\ref{TQ}). 

In \S 3 we review the coordinate Bethe ansatz and show how to relate the wave
function components of the eigenvectors of the $SU(3)$ quantum chain with
periodic boundaries in terms of the components of the eigenvectors of
the inhomogeneous 6-vertex model. In particular all the free-fermion-like energies
are related to a single factorizable eigenvalue of the inhomogeneous 6-vertex
model.

In \S 5 exploring the existence of the free-fermion-like solutions of the
$SU(3)$ chain at $q=\exp{i\frac{2\pi}{3}}$ we show how to produce the recurrence
relations that allows the computation of the wave vector amplitudes related to 
the special factorizable eigenvector of the 6-vertex model. These relations,
although not simple, give us a systematic way to derive all the eigenfunctions 
of the free-fermion part of the eigenspectrum of the periodic $SU(3)$
Perk-Schultz model at $q = \exp{i\frac{2\pi}{3}}$. 

As an application let us consider the free-fermion branch of this last model in
the sector $S_k = (k,k+1,L-2k-1)$. From (\ref{C1}) the corresponding
eigenenergies are given by
 $$
 E_I=-\sum_{j \in I} (1+2 \cos (2\pi j/L)),
 $$
 where $I$ is any  subset of  $(1,2,\ldots,L)$ with  $n=2k+1$ 
 distinct elements. Enumerating these elements by the 
 index $\alpha =1,2,\ldots,n$,
 the  wave function is given by 
  $$
  f(x_1,Q_1;\ldots.x_n,Q_n)=\sum_{P} A^{Q_1,\ldots,Q_n}_{P_1,\ldots,P_n}
  e^{i(k_{P_1} x_1+\ldots+k_{P_n} x_n)},
  $$
  where 
  $k_{\alpha}=2\pi j_{\alpha}/L,  \quad  \alpha =(1,2,\ldots,n)$ 
  are the momenta of the elementary free-fermion excitations.

  Let us  limit ourselves to the  subsets $I$ with elements
  $j_{\alpha}$, satisfying  the constraint 
  $j<m$ or $L-m <j$, where $m$ is a positive integer.
  Due to conjecture 2 of (\cite{AS2})
  the lowest eigenenergy  in the sector $S_k=(k,k+1,L-2k-1)$
  belongs to this part of spectrum if we choose $m>k$.

  Now we fix  $k$ and $m$ with  $m>k$ 
  and consider the bulk  limit  $L 
  \rightarrow 
  \infty$.
   Due to above mentioned constrains for the values of $k$ 
   we have two possibilities:
    $ k_{\alpha} \rightarrow 0$ or $k_{\alpha} 
\rightarrow  2\pi$.
	Consequently from (\ref{Smatrixu}) 
	 all the parameters $u_{\alpha}$ that  fix the auxiliary 6-vertex
	model
	become equal to $\pi/2$ and we obtain the homogeneous model with the
	special
	 eigenvector $A^{Q_1,\ldots,Q_n}$  which (up to  a sign factor) does not
	 depend on the particular 
	 permutation $P$!
	 The wave function can then be written as 
\begin{equation} \label{estre}
	 f(y_1,Q_1;\ldots;y_n,Q_n)  = A^{Q_1,\ldots,Q_n} \sum_{P} (-1)^P
	 e^{2\pi i(j_{P_1} y_1+\cdots+j_{P_n} y_n)},
\end{equation} 
	 where we introduced the  new coordinates   $y_i=x_i/L$ ($i=1,\ldots,n$).
	 In particular this result shows that 
	 in the sectors  ($k,k+1,\infty$) there exist  a lot of eigenstates 
	 (including the one with lowest 
	  eigenenergy in the sector) whose wave functions components  are given  
	 by the  product of  Slater determinants and
	 the components of the ground state wave function of the  
	 XXZ model  with $2k+1$ sites.
	 Let us consider for example the sector $S_2=(2,3,L-5)$. 
	 In order to obtain the lowest eigenenergy in this sector we chose 
	  $I={1,2;L-2,L-1,L}$.  
	 The Slater determinant in (\ref{estre}) reduces to the Vandermonde determinant and we obtain
	 (up
	 to
	 a normalization)
	 \begin{equation}
	 f(y_1,Q_1;\ldots,y_5,Q_5)=\prod_{1 \le j < k \le 5} \sin \pi (y_j-y_k)
	 A^{Q_1,\ldots,Q_5},
	 \end{equation}
	 where we have
	 \begin{eqnarray}
	 &&A^{00111}=A^{01110}=A^{11100}=A^{11001}=A^{10011}=1 \nonumber \\ 
	 &&A^{01011}=A^{10110}=A^{01101}=A^{11010}=A^{10101}=2. \nonumber
	 \end{eqnarray}
	 From these components we see, for example, that the 
	 probability to find 0-particles , separated by
	 1-particles
	 is equal to 
	$ 2*2/(1*1+2*2)=4/5$.

 Moreover, it is important mention that when the inhomogeneity of the
 auxiliary 6-vertex model disappears, and  the wave function  that
 corresponds to the special factorizable eigenvalue is the same as that of 
 the ground state
 of the XXZ spin chain with $\Delta=-1/2$, which  possess for 
 $L=2n+1$ quite 
interesting combinatorial properties.

Due to the conjecture announced at the end of section 2, we have a generalization of this special
wave function to the $SU(N-1)$ - invariant case ($q=e^{i(N-1)\pi/N}$).
So we suspect that the ground state function of the $SU(N-1)$ - invariant spin chain can also
exhibit   interesting combinatorial properties for the special value $q=e^{i(N-1)\pi/N}$. Indeed,
using a bruteforce diagonalization of these quantum invariant chains we have found that
the ratio defined by  
\begin{eqnarray}
\label{Ratio}
R_L=\frac{(\sum_{i} v_i)^2}{\sum_{i} v_i^2}
\end{eqnarray}
where $\{v_i\}$ are the wave function components of the ground state, 
has  a simple form depending on the boundary condition and on the parity of the
lattice size.
For the $L$-sites XXZ spin chain at $\Delta=-1/2$ we have 
$R_L=\sqrt{3}^{\alpha}$ \cite{ALL}, where 
$\alpha=L-1$ or $\alpha=L$ depending if the length $L$ of the chain 
is odd or even, respectively. In the case where $L$ is odd the 
chain has the boundary condition periodic or $SU_q(2)$ invariant, 
and for even values of $L$ the chain has boundary condition 
of twisted type or a $SU_q(2)$ invariant one. 
We can present these results in a compact form:

\begin{eqnarray}
\label{Ratio2}
R_{L+2}/R_{L}=3, \quad R_1=1,\>R_2=3. 
\end{eqnarray}

The numerical results coming from 
 bruteforce numerical diagonalizations 
of chains with $SU_q(N)$ symmetry (free boundary condition) followed by a fitting  with 
special irrational numbers give us  
the values of $R_L$ for the next cases:

\begin{itemize}
\item
 $SU_q(3)$ with $q=-e^{i\pi/4}$ \quad  $2 \le L \le 9$
 \begin{eqnarray}
\label{Ratio3}
&&R_{L+3}/R_{L}=(1+\sqrt{2})^3 \\
&& R_1=1\quad\>R_2=1+\sqrt{2}\quad\>R_3=(1+\sqrt{2})^3. \nonumber
 \end{eqnarray}
\item
$SU_q(4)$ with $q=-e^{i\pi/5}$ \quad   $2 \le L \le 8$      
 \begin{eqnarray}
\label{Ratio4}
&&R_{L+4}/R_{L}= (5+2\sqrt{5})^2 \\
&&R_1=1\quad \>R_2=\sqrt{5}\quad \>R_3=5+2\sqrt{5}\quad \>R_4=(5+2\sqrt{5})^2. \nonumber
\end{eqnarray}
\item
$SU_q(5)$ and $q=-e^{i\pi/6}$ \quad  $2 \le L \le 7$  
 \begin{eqnarray}
\label{Ratio5}
&&R_{L+5}/R_{L}=(2+\sqrt{3})^5 \nonumber \\
&&R_1=1\quad \>R_2=(2+\sqrt{3})/\sqrt{3}\quad \>R_3=
(2+\sqrt{3})^2/\sqrt{3} \\
&&R_4=(2+\sqrt{3})^3\quad\>R_5=(2+\sqrt{3})^5. \nonumber
\end{eqnarray}
\end{itemize}

The numbers  $R_L$, $L<N$ which are necessary for the use  of  the
recursion relations can be found from the explicit expression of  the corresponding wave
function. Using an approach, described in Appendix B of paper\cite{AS2} one can 
show that for $L<N$:
 \begin{eqnarray}
\label{Ratiogen}
&&R_{L}=\biggl ( \frac{1+x}{1-x}\biggr )^{L-1} 
\prod_{k=2}^L\frac{1-x^k}{1+x^k}\quad x=-q=e^{i\pi/N}.
\end{eqnarray}
For the cases where  $L \ge N$ although we cannot prove, our numerical 
results indicate the 
nice recursion relation
\begin{eqnarray}
\label{RR}
&&R_{L+N-1}=R_{N-1}\>R_{L}.
\end{eqnarray}

{\it Acknowledgments}  This work was supported 
in part by the brazilian agencies FAPESP and CNPQ (Brazil), by the Grant 
 \# 01--01--00201 (Russia) and INTAS 00-00561.

\end{document}